\documentclass[prl,showpacs,preprintnumbers,amsmath,amssymb,superscriptaddress,nofootinbib,english,twocolumn]{revtex4}
\usepackage{graphicx}
\usepackage{dcolumn}
\usepackage{bm}
\usepackage{epsfig}
\usepackage{graphicx}
\usepackage{hyperref}
\usepackage[usenames]{color}
\usepackage{url}
\hypersetup{
    colorlinks=true,
    linkcolor=black,
    citecolor=black,
}

\newcommand{\remove}[1]{}

\def\be{\begin{equation}}
\def\ee{\end{equation}}
\def\ba{\begin{eqnarray}}
\def\ea{\end{eqnarray}}

\begin{document}

\title{Lorentz Invariance Violation in Modified Gravity}

\author{Philippe~Brax}
\email[Email address: ]{philippe.brax@cea.fr}
\affiliation{Institut de Physique Theorique, CEA, IPhT, CNRS, URA 2306, F-91191Gif/Yvette Cedex, France}

\date{\today}

\begin{abstract}
We consider an environmentally dependent violation of Lorentz invariance in scalar-tensor models of modified gravity  where General Relativity is retrieved locally thanks to a screening mechanism. We find that  fermions have a modified dispersion relation and would go faster than light in an anisotropic and space-dependent way along the scalar field lines of force. Phenomenologically, these models are tightly restricted by the amount of Cerenkov radiation emitted by the superluminal particles, a constraint which is only satisfied by chameleons. Measuring the speed of neutrinos emitted radially from the surface of the earth and observed on the other side of the earth would probe the scalar field profile of modified gravity models in dense environments.
We argue that the test of the equivalence principle provided by the Lunar ranging experiment implies that a deviation from the speed of light, for natural values of the coupling scale between the scalar field and fermions, would be below detectable levels, unless gravity is modified by camouflaged chameleons where the field normalisation is environmentally dependent.
\end{abstract}

\maketitle
The discovery of the acceleration of the Universe has led to a flurry of scenarios involving
scalar fields and leading to different types of modified gravity models \cite{Clifton:2011jh}. All of them allow for large deviations
from General Relativity on astrophysical scales while preserving Newton's law locally in the solar system and in laboratories on earth.
This is achieved thanks to screening features such as the Vainshtein mechanism for theories with higher order derivative self interactions\cite{vain}, or the chameleon\cite{chameleon}, symmetron\cite{symmetron} or Damour-Polyakov properties\cite{dp1994} for theories with non-linear effective potentials in the presence of pressure-less matter. Recently and after the claim of super-luminal propagation of neutrinos by the OPERA experiment\cite{opera}, it has been suggested that fermions may travel faster than the speed of light in dense environments where the presence of matter offers a breaking of Lorentz invariance\cite{dvali}. This was further pursued in \cite{Kehagias:2011cb,Saridakis:2011eq,Hebecker:2011yh} and then in \cite{Ciu} where Galileons were used to describe the OPERA claims although failing to respect the tight bounds on the deviation of the electron speed from the speed of light. In this work, we will describe a general mechanism which leads to anomalously fast fermions and where  the breaking of Lorentz invariance by a two-tensor appears naturally in models of modified gravity with screening properties.  Chameleon, symmetron and Galileon models can reproduce phenomena of the type indicated by the preliminary OPERA publication. Only models with a thin shell screening mechanism, i.e. chameleons, are compatible with the constraints on the speed of charged leptons. Unfortunately for  chameleons, Galileons and symmetrons and for natural values of the coupling between the scalar field and fermions we find that the deviation of the fermion speed  from the speed of light would be  unobservably small. More generally and model independently,    testing the properties of neutrinos in  matter probes the scalar field profile of modified gravity models and gives us an unprecedented opportunity to analyse the properties of modified gravity in dense environments. We argue that the tests of the equivalence principle such as the Lunar Ranging experiment imply that no deviation from the speed of light due to modified gravity should be observable in the neutrino sector of the standard model unless the scalar field is a camouflaged chameleon. Camouflaged chameleons are such the field normalisation is environmentally dependent and the coupling between the scalar field and fermions becomes smaller in dense media than in vacuum.

We consider the  action governing the dynamics of a scalar field $\phi$ in a scalar-tensor theory
of the general form
\begin{eqnarray}
S &=& \int d^4x\sqrt{-g}\left\{\frac{m_{\rm Pl}^2}{2}{
R}-F(\phi,\partial_\mu\phi,\partial_\mu\partial_\nu \phi)- V(\phi)\right\}\nonumber\\
&& + \sum_i S_m^i (\psi_m^{(i,j)},\tilde g^i_{\mu\nu})\,, \label{action}
\end{eqnarray}
where
\begin{equation}
S_m^i (\psi_m^{(i,j)},\tilde g^i_{\mu\nu})=\int d^4x \sqrt{-\tilde g_i} {\cal
L}^i_m(\psi_m^{(i,j)},\tilde g^i_{\mu\nu})
\end{equation}
is the action in the ith sector of the model,
$g$ is the determinant of the metric $g_{\mu\nu}$, ${ R}$ is
the Ricci scalar and $\psi_m^{(i,j)}$ are various matter fields
labeled by $j$ interacting with the metric $\tilde g_{\mu\nu}^i$ in the Lagrangian ${\cal L}^i_m$.
 When $F(\phi,\partial_\mu\phi,\partial_\mu\partial_\nu \phi)=\frac{1}{2}(\partial\phi)^2$, the field is canonically normalised. More complex functions $F(\partial_\mu\phi,\partial_\mu\partial_\nu \phi)$ appear
in the Galileon scenario for instance \cite{galileon}. A key ingredient of the model is the
coupling of $\phi$ with matter particles. More precisely, the
excitations of each matter field $\psi_m^{(i,j)}$ couple to
a metric $\tilde g_{\mu\nu}^i$ which is related to the
Einstein-frame metric $g_{\mu\nu}$ by
\begin{equation}
\tilde g_{\mu\nu}^i=A^2(\phi)g_{\mu\nu}^i
\end{equation}
where
\begin{equation}
g_{\mu\nu}^i=g_{\mu\nu}+\frac{2\partial_\mu \phi\partial_\nu\phi}{M_i^4}
\end{equation}
depends on each sector of the theory. For instance, the fermion kinetic terms may couple to a different metric from the gauge kinetic terms.

In the following, we will assume that the bi-metric term $\frac{\partial_\mu \phi \partial_\nu\phi }{M_i^4}$ is a small correction to $g_{\mu\nu}$.
In the context of extra dimensional models where matter lives on a brane, the scalar field can be seen as parameterising the normal to the brane, the coupling function $A(\phi)$ arises from the warping of the bulk metric while
the bilinear term reflects the coupling of matter to the induced metric on the brane.
Defining the energy momentum tensor in each sector as
$
T^{\mu\nu}_i=\frac{2}{\sqrt{-g}} \frac{\delta S_m^i}{\delta g_{\mu\nu}}
$
and expanding the action to linear order, we find that the scalar field couples derivatively to matter
\begin{eqnarray}
&&S = \int d^4x\sqrt{-g}\left\{\frac{m_{\rm Pl}^2}{2}{
R}-F(\partial_\mu\phi,\partial_\mu\partial_\nu\phi)- V(\phi) \right\}\nonumber\\
&& + \sum_i \int d^4x\sqrt{-g} \frac{\partial_\mu \phi\partial_\nu \phi}{ M_i^4} T_i^{\mu\nu}\nonumber \\ && + \sum_i \int d^4x \sqrt{- g}A^4(\phi) {\cal
L}_m(\psi_m^{(i,j)},A^2 (\phi) g_{\mu\nu})\,, \label{action}
\end{eqnarray}
As soon as $\theta_{\mu\nu}=\partial_\mu\phi\partial_\nu\phi$ does not vanish due to the presence of matter, we find that Lorentz invariance is broken and
a new Lorentz invariance violating coupling to the matter energy momentum tensors is present in the model\cite{Kehagias:2011cb,Hebecker:2011yh}.

Massless fermions with the action
$
S_F=- \int d^4x \sqrt{-g} \frac{i}{2}(\bar \psi \gamma^\mu D_\mu \psi - (D_\mu \bar \psi) \gamma^\mu \psi)
$
have the energy momentum tensor
$
T^F_{\mu\nu}= -\frac{i}{2} (\bar \psi \gamma_{(\mu} D_{\nu)} \psi - (D_{(\mu }\bar \psi) \gamma_{\nu)} \psi)
$
symmetrised over the indices. This induces the following interaction terms with the scalar field
\begin{equation}
-\frac{i}{2}\int d^4 x \sqrt{-g} \frac{\partial^\mu \phi \partial^\nu\phi}{M_\psi^4} (\bar \psi \gamma_{(\mu} D_{\nu)} \psi - (D_{(\mu }\bar \psi) \gamma_{\nu)} \psi)
\end{equation}
where the mass scale $M_\psi$ is the suppression scale for the fermion species $\psi$. A similar coupling was first considered in \cite{gauthier}.

Let us focus on a typical situation where the metric is Minkowskian to a good approximation (this is the case on earth where Newton's potential is $\Phi_\oplus \sim 10^{-9}$) and the scalar field varies on scales much larger than the Compton wave length of the fermions. Moreover, let us assume that the scalar field is static. Then the interaction term reduces to
\begin{equation}
-\frac{i}{2}\int d^4 x \sqrt{-g} \  d^i d^j (\bar \psi \gamma_{i} \partial_j \psi- \partial_i \bar \psi \gamma_j \psi)
\end{equation}
where
$
d^i= \frac{\partial^i\phi}{M_\psi^2}
$
is a slowly varying function of space only.
Hence a static configuration of the scalar field yields a Lorentz violating interaction in the Fermion Lagrangian.
The resulting Dirac equation becomes
\begin{equation}
i(-\gamma^0\partial_0 +\gamma^i \partial_i +d^i d^j \gamma_i \partial_j)\psi=0.
\end{equation}
The dispersion relation is obtained by squaring the modified Dirac operator to obtain
\begin{equation}
p_0^2 = (c^2)^{ia}p_i p_a.
\end{equation}
This becomes the dispersion relation in an anisotropic medium with a square velocity tensor
\begin{equation}
(c^2)^{ia}=(\delta^{ij}+ d^i d^j)(\delta^{aj} + d^a d^j).
\end{equation}
The eigenmodes of the velocity tensor are $d^i$ and two vectors $e^i_\lambda, \lambda=1,2$ orthogonal to $d^i$.
The eigenspeeds are $c_d=(1+\vert d\vert^2)$ and twice $c_\lambda=1$. Hence, fermions go  faster than light in the direction of the
gradient $\partial^i \phi$, i.e. along the scalar lines of force,  with
\begin{equation}
\Delta c\equiv c_d-1= \vert d\vert^2.
\end{equation}
An interesting application concerns fermions, typically neutrinos with no electromagnetic interactions,  produced at the surface of a spherical body, traversing the sphere on a straight line parameterised by an angle $\theta$ varying between $-\theta_{\rm max}$ and $\theta_{\rm max}$. We assume that $d^i$ is radial. Along this line, the fermion speed is
$
v(\theta)=1 + \vert d\vert^2 \theta^2
$
implying that fermions would be observed earlier compared to a propagation with the speed of light by
\begin{equation}
\frac{\Delta t}{t}= \frac{1}{3} \vert d\vert^2 \theta_{\rm max}^2.
\end{equation}
A time difference of $\frac{\Delta t}{t}=2.5\ 10^{-5}$, as claimed by the OPERA experiment where $\theta_{\rm max}\sim 0.06$, would require $\vert d\vert \sim 0.15$. In the following we will use such a value as a template and unravel its physical consequences. If the OPERA claims turned out to be spurious, the mechanisms described
here would still be valid and lower bounds on the scale $M_\psi$ would ensue. These bounds would be easily extracted from the formalism discussed here.

When the scalar field is canonically normalised,
the Klein Gordon equation is modified due to the coupling of the scalar field $\phi$ to matter:
\begin{equation}
\Box \phi- \sum_i \frac{2}{M_i^4} D_\mu (\partial_\nu \phi T_i^{\mu\nu})= -\beta \sum_i T_i + \frac{dV}{d\phi},
\end{equation}
where $T_i$ is  the trace of the energy momentum tensor $T_i^{\mu\nu}$ and
the coupling of $\phi$ to matter is defined by
$
\beta\equiv m_{\rm Pl}\frac{d\ln A}{d \phi}.
$
In static situations where matter is pressure-less and space-time is assumed to be Minkowskian, the Klein-Gordon equation reduces to
\begin{equation}
\Delta \phi = -\beta  T + \frac{dV}{d\phi}
\end{equation}
corresponding to the case with no derivative coupling in the Lagrangian.

In a dense environment, the scalar field acquires a non-trivial profile due to the matter dependent source term.
We will be interested in spherical situations corresponding to dense astronomical or astrophysical objets such as the earth or the sun.
In the superluminal context, this was first considered in \cite{Kehagias:2011cb} where the role of scalar fields coupled to fermions was emphasized. In the absence of a potential term and assuming that the coupling to matter is also determined by the scale $M_\psi$, i.e. we have $\beta=\frac{m_{\rm Pl}}{M_\psi}$, a time difference of order $\frac{\Delta t}{t}=2.5\ 10^{-5}$ can be explained if $M_\psi\approx 1\ {\rm TeV}$. This scale is large enough to argue that the coupling of the scalar field to matter may come from integrating out heavy fields at a scale which lies beyond the standard model of particle physics. Unfortunately, it leads to a very large value of the coupling to matter $\beta$ which would be ruled out by experimental tests of Newton's law.
This can be avoided in modified gravity models where the effects of the scalar field are screened in local tests of gravity. This is the type of models we investigate in this paper. We will first focus on a large class of known models, i.e. chameleons, Galileons and symmetrons. We will then  deal with more model independent considerations and eventually introduce camouflaged chameleons where chameleon fields have an environmentally dependent normalisation.

Large deviations from Newton's law are prevented in chameleon-like theories\cite{chameleon} where the effective
$
V_{\rm eff}(\phi) = V(\phi) +\rho_m A(\phi)
$
acquires an environment dependent minimum where the mass is large enough to Yukawa screen the fifth force deep in the body, leaving only a thin shell of size $\Delta R$ over which the field varies significantly. Here we have defined $\rho_m$ as the conserved matter density $T=-A(\phi)\rho_m$. In summary, for radial distances $r\le R_s$ where $R_s$ is the radius of the shell, the solution is constant
$
\phi=\phi_c, \ \ r\le R_s
$
where $\phi_c$ is the minimum of the effective potential inside the dense body and we assume $A(\phi)\approx 1$. In the thin shell, the field varies according to
$
\phi=\phi_\infty -\frac{\beta \rho_m R^2}{2 m_{\rm Pl}} +\frac{\beta \rho_m R_s^3}{3 m_{\rm Pl} r} + \frac{\beta\rho_m r^2}{6m_{\rm Pl}}.
$
Outside the body, the fifth force is suppressed by a factor
$
\frac{3\Delta R}{R} \equiv \frac{ \vert \phi_\infty -\phi_c\vert}{2\beta m_{\rm Pl}\Phi_N}
$
where $\Phi_N$ is the Newton potential generated by the body at its surface and $\Delta R= R-R_s$.
A thin shell exists when $3 \Delta R / R \lesssim 1$.
Solar system tests of gravity like the Lunar Ranging experiment require that the thin shell on earth should be such that  $\beta \frac{\Delta R_\oplus}{R_\oplus}\lesssim 10^{-7}$ \cite{chameleon}.

Inside a dense body, the scalar field has a non-vanishing gradient in a thin shell.
More precisely we find that the gradient of the scalar field is radial
$
\partial^i \phi =\frac{d\phi}{dr} \frac{x^i}{r}
$
where
$
\frac{d\phi}{dr}= -\frac{\beta \rho_m R_s^3}{3 m_{\rm Pl} r^2} + \frac{\beta\rho_m r}{3m_{\rm Pl}}, R_s \le r\le R,
$
 vanishes for $r\le R_s$ and can be approximated by
$
\frac{d\phi}{dr}= \frac{\beta \rho_m }{ m_{\rm Pl} }(r-R_s), R_s\le r\le R,
$
which is maximal at the surface of the body and proportional to $\Delta R$.
As an example, let us consider fermions starting from the surface of the dense body and traversing the body at a small angle $\theta$ from the horizontal which starts at $-\theta_{\rm max}$ and finishes at $\theta_{\rm max}$. Define by $\theta_{\rm min}$ the angle when the fermions leave the thin shell. We have
$
\frac{d\phi}{dr}= \frac{\beta \rho_m R}{ 2 m_{\rm pl}} (\theta^2 -\theta_{\rm min}^2)
$
for small angles and
$
\Delta \theta= \frac{\Delta R}{R\theta_{\rm max}}
$
with $\Delta \theta= \theta_{\rm max} -\theta_{\rm min}$.
Along this trajectory the speed of the fermions is
$
v(\theta)= 1+ \vert d\vert^2 \theta^2
$
corresponding to the difference of time of arrival between fermions and photons
$
\Delta t= R\int_{\theta_{\rm min}}^{\theta_{\rm max}} (1-\frac{d\theta}{1+\vert d^2 \vert \theta^2}).
$
Denoting by $t= R\theta_{\rm max}$ the time photons take to go from $\theta_{\rm max}\sim \theta_{\rm min}$ to $\theta=0$, we find that
\begin{equation}
\frac{\Delta t}{t}= \frac{12 \Phi_N^2 m_{\rm Pl}^2}{\beta R^2 M_\psi^4} (\beta \frac{\Delta R}{R})^3.
\label{tt}
\end{equation}
As the thin shell is very small, this time difference can reproduce $\frac{\Delta t}{t}\sim 2.5 \ 10^{-5}$ with a low suppression scale  $M_\psi \lesssim 10^{-1}$ eV. The smallness of this scale can be understood as resulting from the large suppression of the scalar field fifth force by the thin-shell mechanism. As it stands, such a low value of $M_\psi$ appears to be much too low to be justifiable in an effective field theory context where we  understand the origin of the coupling between the scalar field and fermions as a result of integrating out heavy fields. This points out that a result of the OPERA type would not be explained in a natural way by chameleon models. We will introduce camouflaged chameleons to deal with this issue in a positive way.

Some models of modified gravity can be compatible with gravitational tests while describing scalar fields as being unscreened on earth. This is the case of the symmetron where the potential and the coupling functions are
$
V(\phi)= -\frac{\mu^2}{2} \phi^2 +\frac{\lambda}{4}\phi^4,\ A(\phi)= 1 + \frac{\phi^2}{2M_G^2}.
$
For a low energy density, the field is stabilised at a value $\phi_\star=\frac{\mu}{\sqrt{\lambda}}$ whilst for a large energy density $\rho\ge \rho_\star=\mu^2 M_G^2$, the minimum of the effective potential is at the origin.
The solution inside an unscreened body reads
$
\phi(r)= A\frac{R}{r} \sinh \frac{\sqrt{\rho_m}}{M_G} r
$
where $A\sim \frac{M_G^3}{m_{\rm Pl}^2} \frac{1}{\sqrt{6\Phi_N}}$ with $M_G\le 10^{-3} m_{\rm Pl}$ to comply with solar system tests.
In particular we have
$
 \frac{d\phi}{dr}\vert_{r=R}= {\cal O}(\frac{A}{R})
$. We will focus on the upper value of $M_G$  which corresponds to $\phi_\star \sim 10^{-6} m_{\rm Pl}$.
 A  contribution of order $\vert d\vert \sim 0.15$ could be explained by these models provided $M_\psi^2 \sim 10 A/R$, or equivalently $M_\psi\sim 7\cdot 10^{-5}$ GeV. Again, such a low value appears to be too low to be justifiable from an effective field theory point of view.

Galileon models \cite{galileon}   depend on four unknown parameters and involve three non-canonical contributions to the kinetic terms. We consider the case with $c_{4,5}=0$ for simplicity. Inside the Vainshtein radius expressed as
$
R_\star= (4 c_3 \Phi_N \frac{R}{m_{\rm Pl}^2})^{1/3}
$
the Galileon profile due to a spherical mass $M$ reads
$
\frac{d\phi}{dr}= \frac{M}{2m_{\rm Pl}r^2}(\frac{r}{R_\star})^{3/2}
$
where we have normalised the Galileon choosing $c_2=m_{\rm Pl}^2$.
In this case, we find that the gradient of the scalar field is
$
\frac{d\phi}{dr}\vert_R=  m_{\rm Pl}^2 (\frac{\Phi_N}{4c_3})^{1/3}.
$
Taking $c_3 \sim 10^{120}$\cite{Burrage:2010rs,brax2} to satisfy the Lunar Ranging tests, we find that $d\sim 0.15$ could be due to a galileon coupled to matter
provided
$
M_\psi\sim 1.3\ {\rm MeV}.
$
Although very low, such a value may be more justifiable in the Galileon context as non-renormalisation theorems exist \cite{Hui:2010dn}. Moreover, the Galileon normalisation varies with the environment and the coupling scale becomes much larger close to dense media where the coupling scales to matter, viewed as the cut-off of the theory, become of natural values.

If the coupling to $g^i_{\mu\nu}$ respects gauge invariance and if flavour effects are taken into account \cite{cohen,strumia}, all neutrinos of the standard model should couple with the same scale $M_\psi$. This would also entail that electrons and muons would have the same speed as the neutrinos at the classical level inside the earth. In the atmosphere there is a drastic difference between models with a thin shell effect like chameleons for which $\phi \sim \phi_{\rm atm}$ sits at the minimum of the effective potential \cite{chameleon} and induces no change in the speed of fermions at all  and models with no thin shell like symmetrons and Galileons  where $d$ is nearly constant in the atmosphere. The latter would strongly violate the bounds on the deviation from the speed of light for electrons  as they are in the $10^{-15}$ range\cite{strumia}. One possibility for these models would be to have an environmentally dependent violation of gauge invariance with $M_\nu\ne M_f$ for $f=e,\mu,\tau$. A large value of $M_f\ge 10^4 M_\nu$ would be enough to satisfy the experimental bounds. In vacuum where the scalar field has no gradient, gauge invariance would be retrieved. Unfortunately, the resulting large difference between the neutrino and the electron speeds inside the earth would lead to too much Cerenkov radiation and is therefore excluded\cite{cohen}.

The only viable models are the chameleon ones where $d$ vanishes in the atmosphere and in the vacuum pipes of particle experiments, implying no deviation  between the fermion speed and the speed of light there. Moreover, the fact that neutrinos are obtained from pion decay in vacuum where the neutrino speed is the speed of light implies that no bound on the resulting neutrino energy applies \cite{Bi:2011nd}. Similarly, in matter such as inside the earth, as long as all fermions couple to the same metric with the same $M_\psi$, no $e^+e^-$ Cerenkov radiation is induced.
For chameleon models, Cerenkov photon radiation $\nu\to \nu +\gamma$ happens in the thin shell only and leads to an energy loss \be \frac{\delta E}{E}\approx -\frac{k}{84 \pi^4}\frac{\alpha G_F^2 \theta_{\rm max}^5E^5 \rho_m^6 R_\oplus^7 }{\beta  m_{\rm Pl}^6 M_\psi^{12}} (\frac{\beta \Delta R_\oplus}{R_\oplus})^7 \ee where $k=25/448$ which can be a tiny effect if the size of the thin shell on earth is extremely small.

So far we have taken $\frac{\Delta t}{t}\sim 2.5 \ 10^{-5}$. If the OPERA claims were not confirmed and $\frac{\Delta t}{t}$ happened to be smaller but still  greater than the bound
on the deviation of the speed of charged leptons from the speed of light, i.e. $\frac{\Delta t}{t} \gtrsim 10^{-15}$, modified gravity models with no thin shell such as Galileons and symmetrons would not provide an explanation for such a phenomenon. Indeed, the deviation from the speed of light for neutrinos and charged leptons would be the same both inside the earth and in the atmosphere, implying a direct violation of the bound on the deviation of the speed of charged leptons from the speed of light. On the other hand, chameleons would still be able to reproduce a value of $\frac{\Delta t}{t}\gtrsim 10^{-15}$ with no violation of the bound on the speed of charged leptons deduced from experiments in particle accelerator vacua and the earth's atmosphere. Indeed, the gradient of the chameleon field vanishes inside these environments. As a result, this would simply imply that $M_\psi$ would be larger than a few eV's. Eventually, if experiments  concluded that no significant deviation from the speed of light for neutrinos and charged leptons could be detected, this could simply mean that $M_\psi$ is large enough to lead to no experimentally detectable effects.  Moreover, it is to be expected that $M_\psi$ should be larger than the electroweak scale if the coupling between matter and the scalar field results from the dynamics of theories beyond the standard model. In this case, the time difference $\frac{\Delta t}{t}$ would be unobservably low. We will see that this negative conclusion can be avoided if the scalar field is a camouflaged chameleon.

In the chameleon,  symmetron and  Galileon cases, we have found that a compatibility with $\frac{\Delta t}{t}\sim 2.5 \ 10^{-5}$ could be reached for low values of $M_\psi$. As already argued, these scales are unnaturally low. In order to assess more quantitatively how unnatural these scales are, we now analyse the large effects they may have in particle physics experiments such as the ones at LEP, i.e. increasing the width of the Z boson or
modifying the electroweak precision tests. To do so, we focus on the coupling of the scalar field to the gauge fields of the standard model and we assume  a low scale $M_F\sim M_\psi$.  Although a detailed study is beyond the present work, a simplified analysis can be carried out.  One can evaluate the order of magnitude of such effects by reducing the
bilinear coupling of the scalar field to W and Z bosons, to a linear coupling. First of all notice that the energy momentum of gauge fields involves
$
T^F_{\mu\nu} \supset \frac{1}{4} g_{\mu\nu} F^2
$
leading to the effective coupling between the scalar field and the gauge bosons
$
{\cal L}_I=-\frac{1}{4M_F^4} (\partial \phi)^2 F^2.
$
In the vacuum where particle physics experiments take place, in the chameleon and symmetron cases, and expanding $\phi=\phi_{\rm vac} + \delta\phi$, this leads to the operator, after one integration by parts,
$
{\cal L}_I \supset \frac{\phi_{\rm vac}}{4M_F^4} \partial^2 \delta \phi F^2.
$
The gauge boson vacuum polarisation diagrams receive contributions from scalar loops. The effect of these loop is to induce potentially divergent
contributions to the precision parameters $S$, $T$, etc \dots. Fortunately, at high momentum the electroweak symmetry breaking is irrelevant, implying a cancellation
of the UV divergences. This is the essence of the screening theorem for scalars. As a result, only momenta up to the breaking scale $M_Z$ are relevant. This implies that we can replace the previous operator by
$
{\cal L}_I \sim \frac{\phi_{\rm vac} M^2_Z}{4 M_F^4} \delta \phi F^2.
$
The effect of such a vertex was studied in \cite{part} where the suppressions scale
$
\hat M_F = \frac{4M_F^4}{\phi_{\rm vac} M_Z^2}
$
was constrained to be $\hat M_F \ge 1 \ {\rm TeV}$. Here we find that $\hat M_F \sim 10^{-35} {\rm GeV}$ for $M_F\sim M_\psi$ in the symmetron case where $\phi_{\rm vac}=\phi_\star$. Of course this is  a situation which is strongly excluded as the effects on the precision tests of the standard model would be dramatic. On the other hand, if we take into account the bound $\hat M_F \ge 1$ TeV we find that the precision test bound is satisfied provided $M_F \ge 14 \ {\rm TeV}$. Assuming that the couplings of the symmetron to both matter and gauge fields have a similar origin, this would certainly indicate that $M_\psi$ should be  larger than a few TeV's.

For chameleons\cite{chameleon}, $\phi_{\rm vac}\le 10^{-28} m_{\rm Pl}$, implying that $ M_F \ge 0.15\ {\rm GeV}$. Hence a value of $M_F$ slightly larger than the electroweak scale would be compatible with the standard model precision tests.
In the Galileon case, the Lorentz invariant breaking background leads to the operator
$
{\cal L}_I= -\frac{dM_\psi^2}{4M_F^4}{\partial_r \delta \phi} F^2
$
where $d\sim 0.15$,
which would lead to the same effect in precision tests as
$
{\cal L}_I \sim -\frac{dM_Z M_\psi^2}{4 M_F^4} \delta \phi F^2
$
corresponding to a scale
$
\hat M_F= \frac{4M^4_F}{d M_Z M_\psi^2}
$
which is $\hat M_F \sim 10^{-6} {\rm GeV}$ when $M_F\sim M_\psi$, a value which is much too small. The precision test bound is satisfied provided $M_F\ge 100\ {\rm MeV}$.

All in all, we have found that if the coupling to  matter and to photons have similar origins and share similar scales $M_F\sim M_\psi$, the precision tests of the standard model exclude the low values of the coupling scale $M_\psi$ which would be necessary to reproduce a time difference $\frac{\Delta t}{t}=2.5\ 10^{-5}$ for chameleon models. On the other hand, nothing seems to exclude values of $M_\psi\sim M_F$ slightly beyond the standard model. The coupling to photons with such a strength would certainly lead to strong effects in astrophysics and the laboratory along the lines of \cite{Brax:2007hi}. Work on this topic is in progress\cite{us}.

Having so far focused on known models of modified gravity, i.e. chameleons, Galileons and symmetrons, let us now come back to model independent features. If scalar fields modifying the laws of gravity exist, then their profile in the presence of matter breaks Lorentz invariance. This
Lorentz invariance violation could then be transmitted to  the neutrino sector of the standard model as we have seen in (\ref{action}). In these models, as the variation of the neutrino speed follows the scalar lines of force, the deviation from the speed of light  would be maximal for neutrinos emitted radially towards the centre of the earth and detected in a laboratory symmetrical from the emission point. In this case, the time difference would be
\be
\frac{\Delta t}{t}= \frac{1}{R} \int_0^R dr d^2 (r).
\ee
which depends on the entire profile of the scalar field inside the earth.
For a generic modified gravity model, gravity tests apply outside dense objects where constraints on the scalar field gradient are tight. Inside a dense object, no such constraints are known. It is simply expected that the scalar field gradient will be screened compared to the unscreened case.
In the unscreened case, we have essentially
\be
\frac{d\phi}{dr}= \beta \frac{\rho_m}{3 m_{\rm Pl}}r
\ee
which gives an upper bound for the time difference
\be
\frac{\Delta t}{t}\lesssim \frac{4}{3} \beta^2 \Phi_N^2 \frac{m_{\rm Pl}^2}{M_\psi^4 R^2}.
\ee
For $\beta={\cal O}(1)$, and $M_\psi={\cal O}(1)$ TeV, this upper bounds implies that $\frac{\Delta t}{t}\lesssim 10^{-38}$. Hence we do not expect that any time difference may ever be detected if the bare coupling $\beta$ is of order one. A larger time difference may be obtained if the modified gravity models have a large bare coupling $\beta=\frac{m_{\rm Pl}}{M_\psi}$ corresponding to a universal suppression scale $M_\psi$ for both the derivative and non-derivative couplings of the scalar field $\phi$ to matter. In this case the upper bound becomes
\be
\frac{\Delta t}{t}\lesssim \frac{4}{3}  \Phi_N^2 \frac{m_{\rm Pl}^4}{M_\psi^6 R^2}
\ee
which is of order $2\cdot 10^{-5}$ for the natural scale $M_\psi^2 \sim 10^5\ {\rm GeV}^2$ corresponding to the results in \cite{Kehagias:2011cb}. For this value of the coupling, a deviation from the speed of light greater than the one in the charged lepton sector, i.e. $\frac{\Delta t}{t}\gtrsim 10^{-15}$, can only be achieved provided the screening factor\footnote{For chameleon models, the screening factor is at most of order $s(R)\sim \frac{\Delta R}{R}$ varying over a thin shell of width $\Delta R$ and vanishing otherwise. Here we envisage models where $s(r)$ varies smoothly across the body and is non-vanishing over a finite interval of the order of the size of the dense body.}
\be
s(r) = \frac{\frac{d\phi}{dr}}{ \beta \frac{\rho_m}{3 m_{\rm Pl}}r}
\ee
is in the range $10^{-5}\lesssim s(r) \lesssim 1$. Hence we have found that deviations from the speed of light could be present for neutrinos going through the earth if gravity were modified in a dense environment and two conditions were met. The first one is that the coupling to matter $\beta$ must be large, a situation which was investigated for chameleon models in \cite{Mota:2006ed} for instance. Secondly, the screening of the scalar field profile inside the earth cannot be too large, certainly not at a level below $10^{-5}$.

Of course, a model of modified gravity satisfying these requirements would also have to explain the absence of modified gravity effects in the Lunar ranging experiment at the $\eta_\oplus= 10^{-13}$ level \cite{Williams:2012nc}.  This requires to know the profile of the scalar field outside the dense object.
Assuming that the scalar field behaves like a free field close to the body, we find
\be
\frac{d\phi}{dr} \approx 2 \beta m_{\rm Pl} \Phi_N s(R) \frac{R}{r^2}.
\ee
Notice that when the field is extremely screened inside the dense body $s(R)\approx 0$ implying that the field is nearly constant outside the body and therefore reducing the amount of Cerenkov radiation there. The coupling between dense objects and the scalar field is  of order $\beta s(R)$
implying that the acceleration difference between the moon and the earth in the gravitational field of the sun is of order\cite{chameleon}
\be
\eta_\oplus \approx (\beta s(R))^2.
\ee
The Lunar Ranging experiment leads to
$\beta s(R) \lesssim 10^{-7}$. For large values of $\beta$, this essentially rules out the possibility of having $10^{-5}\lesssim s(R)\lesssim 1$ and therefore the order of magnitude estimate
\be
\frac{\Delta t}{t}\lesssim  \frac{4}{3} \eta \Phi_N^2 \frac{m_{\rm Pl}^2}{M_\psi^4 R^2}
\ee
is unobservably small for natural values of $M_\psi$. Of course, we have assumed here that the field is almost free outside the body.
This is not the case for models of the Galileon type where the field profile is more complex. Nonetheless it seems difficult to reconcile these models with the Cerenkov constraints. This negative result can be altered if the field normalisation becomes environmentally dependent.
This is one of the features of Galileon models where  the coupling scales
to matter for the canonically normalised Galileon become larger around dense media than in vacuum. This mechanism can be transposed to the chameleon models, enabling one to tackle the naturality of the scale $M_\psi$.

Let us extend chameleon models by replacing the canonical kinetic terms by $-\frac{f(\phi)}{2} (\partial \phi)^2$ where $f(\phi)$ is a smooth function. The Klein-Gordon becomes
\be
\Box \phi= \frac{1}{f(\phi)} \frac{dV_{\rm eff}}{d\phi} -\frac{1}{2} \frac{ d\ln f(\phi)}{d\phi} (\partial \phi)^2.
\ee
Vacuum configurations  of chameleon models in a medium of density $\rho_m$ are not modified by the change of normalisation of $\phi$ and are still minima
of the effective potential $V_{\rm eff}$. Denoting by $\phi_c$ the minimum in a medium of density $\rho_c$, the Lagrangian can be linearised using $\phi= \phi_c +\delta \phi$, and the canonically normalised excitation is $\Phi= \sqrt{f(\phi_c)} \delta\phi$. The derivative coupling of this field to matter becomes
\be
\sum_i \int d^4x\sqrt{-g} \frac{\partial_\mu \Phi\partial_\nu \Phi}{ f(\phi_c)M_i^4} T_i^{\mu\nu}
\ee
corresponding to the environmentally dependent coupling scale
\be
M_i(\rho_c)= f^{1/4}(\phi_c)M_i
\ee
while the non-derivative coupling becomes
\be
\beta_i(\rho_c)=\frac{\beta}{f^{1/2}(\phi_c)}
\ee
Let us now study how this change of normalisation affects the thin-shell mechanism and the speed of fermions.

Consider a spherical object of radius $R$ and density $\rho_c$ embedded in a vacuum region of density $\rho_\infty$. Let us assume that this object has a thin shell. For $r\le R_s$, the field is still constant $\phi=\phi_c$. In the thin-shell, we have
\be
\Delta \phi\approx \frac{\beta \rho_m}{f(\phi_c) m_{\rm Pl}}
\ee
corresponding to $\frac{d\phi}{dr}\approx \frac{\beta \rho_m}{f(\phi_c) m_{\rm Pl}} (r-R_s)$ in the shell. Outside the body we have
\be
\phi= \phi_\infty + \frac{(\phi_c -\phi_\infty)R}{r}
\ee
implying that the size of the shell is
\be
\frac{\Delta R}{R}= f(\phi_c) \frac{\phi_\infty -\phi_c}{6m_{\rm Pl} \beta \Phi_N}
\ee
where a factor $f(\phi_c)$ has now appeared compared to the usual chameleon result.

Particles with no thin shell evolve in the background
of a spherical  object by following the trajectories defined by the modified Newtonian potential $\Psi =  \frac{\Phi_N R}{r} + \beta \frac{\phi}{m_{\rm Pl}}$ where $ \frac{\Phi_N R}{r}$ is the newtonian potential at the distance $r$. This is a direct consequence of the fact that particles couple to the metric $A^2(\phi)g_{\mu\nu}$.
If the spherical object has no thin shell and behaves like a point-like particle we have
\be
\Psi=  \frac{\Phi_N R}{r} (1 +2 \beta_\infty^2)
\ee
where $\beta_\infty = \frac{\beta}{f^{1/2}(\phi_\infty)}$ is the coupling of the canonically normalised field $\Phi$ to matter outside the sphere. When the spherical object has a thin shell then
we have
\be
\Psi=  \frac{\Phi_NR}{r} (1 +  2 \beta_\infty \beta_{c,eff})
\ee
where the coupling of the dense object to the scalar field $\beta _{c,eff}$ is obtained to be
\be
\beta_{c,eff}= \frac{f^{1/2}(\phi_\infty)(\phi_\infty -\phi_c)}{2m_{\rm Pl} \Phi_N}
\ee
Notice that we have
\be
\beta_{c,eff}= (\frac{M_\psi(\rho_\infty)}{M_\psi (\rho_c)})^4 3\beta_\infty\frac{\Delta R}{R}=(\frac{M_\psi(\rho_\infty)}{M_\psi (\rho_c)})^2 3\beta_c\frac{\Delta R}{R}.
\ee
As in \cite{chameleon}, the Lunar Ranging constraint depends on $\eta_\oplus  \approx \beta_{c,eff}^2$, implying that $\beta_{c,eff} \lesssim 10^{-7}$ in the solar system.
It is then easy to deduce the time difference $\Delta t$ for fermions through the earth
\be
\frac{\Delta t}{t}= \frac{12 \beta_c^2 \Phi_N^2 m_{\rm Pl}^2}{R_\oplus^2 M^4_\psi (\rho_c)} (\frac{\Delta R_\oplus}{R_\oplus})^3
\ee
which can be expressed as
\be
\frac{\Delta t}{t}= \frac{4}{9}(\frac{M_\psi(\rho_c)}{M_\psi(\rho_\infty)})^4 \frac{\Phi_N^2 m_{\rm Pl}^2}{R_\oplus^2 \beta_\infty M^4_\psi(\rho_\infty)} \eta_\oplus ^{3/2}.
\ee
A natural scale for the coupling scale $M_\psi (\rho_c)$ in matter can be obtained in the strong coupling regime like in \cite{Kehagias:2011cb}
where $\beta_c=\frac{m_{\rm Pl}}{M_\psi (\rho_c)}$. Choosing a reasonable value for the thin shell $\frac{\Delta R}{R} \sim 10^{-1}$, we find that
$\frac{\Delta t}{t} \sim 2\cdot 10^{-5}$ for a coupling scale $M_\psi (\rho_c)\sim 140 \ {\rm GeV}$. The Lunar Ranging constraint can be satisfied with $\eta_\oplus \sim 10^{-7}$ and a low vacuum coupling scale $M_\psi (\rho_\infty) \sim 20 \ {\rm eV}$. As expected, the effect of the camouflage mechanism is to enhance the coupling scale in matter. In vacuum, the coupling scale is still low. Viewed as an order of magnitude for the cut-off of  the effective theory describing the modification of gravity, the coupling scale in matter needs to be large enough to incorporate effects of the standard model. As such a coupling scale like $M_\psi (\rho_c)\sim 140 \ {\rm GeV}$ satisfies this criterion. On the other hand, the effective field theory in vacuum captures effects on very large cosmological scales at energies below the electron mass for which a low cut-off in the ballpark of $M_\psi (\rho_\infty) \sim 20 \ {\rm eV}$ is not contradictory. Of course, more work should be devoted to these camouflaged chameleons. For instance, the variety of coupling scales in different environments
require an appropriate choice
of $f_c$ and $f_\infty$ which have to be very different. This could  be achieved  using power laws for $f(\phi)$ as $\phi_\infty \gg \phi_c$ for most chameleon models.  The phenomenology of these models will be studied elsewhere.

We still need to check that the neutrinos emerging from the supernova SN1987A are not too much in advance compared  to photons. Let us model the trajectory
of the neutrinos as radial from earth, on the verge of the milky way at a distance $R_{\rm E}\sim 8\ {\rm kpc}$ from the galactic centre, to a distance $R_{\rm SN}\sim 60\ \rm {kpc}$. This is not the exact trajectory although it will give us the order of magnitude of the deviation.
Along this trajectory, the neutrino speed varies as $c_d=1 +\vert d\vert^2$ where $d= M^{-2}_\psi\frac{d\phi}{dr}$. The time difference between neutrinos and photons can be easily evaluated in the camouflaged chameleon case, we find that
\begin{equation}
\frac{\Delta t}{t}= \frac{4}{3}\beta_{G,eff}^2 \Phi_N^2 \frac{m_{\rm pl}^2 R_{\rm E}^2}{R_{\rm SN}M_\psi(\rho_\infty)^4}(\frac{1}{R_{\rm E}^3}-\frac{1}{R_{\rm SN}^3})
\end{equation}
where $\Phi_N\sim 10^{-6}$ and  $\beta_{G,eff}\sim 10^{-1}$ for the Milky Way. Here we obtain an  upper bound on the time delay which is typically  $\frac{\Delta t}{t} \lesssim 10^{-16}$.
Therefore,  camouflaged chameleons could provide an explanation for the non-observation of a deviation from the speed of light for neutrinos
emerging from SN1987A. Finally we also have to check that the amount of Cerenkov radiation is small. In fact, this is given by
\be \frac{\delta E}{E}\approx -\frac{k}{84 \pi^4}\frac{\alpha G_F^2 \theta_{\rm max}^5 E^5 \beta_c^6\rho_m^6 R_\oplus^7 }{m_{\rm Pl}^6 M_\psi(\rho_c)^{12}} (\frac{\Delta R}{R})^7 \ee
which is always  small, of order $6\cdot 10^{-4}$.

In conclusion, we have shown that modified gravity models with screening properties can induce Lorentz violation effects in the fermionic sector
of the standard model. Such violations are induced by the profile of a scalar field coupled to matter in the presence of pressure-less over densities.
In particular, fermions have a space-dependent speed along the scalar lines of force. Gravitational and Cerenkov radiation constraints are too tight to expect an observable signal on earth for natural values of the coupling scale between the scalar field and fermions unless gravity is modified by a camouflaged chameleon. Exploring the possible coupling of standard model particles with such scalar fields modifying the laws of gravity is an exciting prospect, still in its infancy, and worth pursuing. New experimental results in the near future will certainly give indications on the possible existence of modifications of gravity by screened scalar fields.

I would like to thank Clare Burrage, Anne Davis and Alexander Vikman for very stimulating  suggestions.

\end{document}